\documentclass[%
 reprint,
 amsmath,amssymb,
 aps,
 prd
]{revtex4-2}

\usepackage{graphicx}
\usepackage{float}
\usepackage{physics}
\usepackage[dvipsnames]{xcolor}

\usepackage[normalem]{ulem}

\newcommand{\szabolcs}[1]{\bgroup\color{magenta} #1\egroup}

\begin{document}

\title{Can rooted staggered fermions describe nonzero baryon density at low temperatures?}

\author{Szabolcs Bors\'anyi$^a$, Zolt\'an Fodor $^{a,b,c,d,e}$, Matteo Giordano$^c$, Jana N. Guenther$^a$, S\'andor D. Katz$^c$, Attila P\'asztor$^{c\dagger}$, Chik Him Wong $^a$}
\address{
$^a$ Department of Physics, Wuppertal University, Gaussstr.  20, D-42119, Wuppertal, Germany \\
$^b$ Pennsylvania State University, Department of Physics, State College, PA 16801, USA \\ 
$^c$ Institute  for Theoretical Physics, ELTE E\"otv\"os Lor\' and University, P\'azm\'any P. s\'et\'any 1/A, H-1117 Budapest, Hungary \\
$^d$ J\"ulich Supercomputing Centre, Forschungszentrum J\"ulich, D-52425 J\"ulich, Germany \\
$^e$ Physics Department, UCSD, San Diego, CA 92093, USA \\
}
\email{$^\dagger$ Corresponding author: apasztor@bodri.elte.hu}

\date{\today}

\begin{abstract}
Research on the QCD phase diagram with lattice field theory methods
is dominated by the use of rooted staggered fermions, as
they are the computationally cheapest discretization available.
We show that rooted staggered fermions at a nonzero baryochemical 
potential $\mu_B$ predict a sharp rise in the baryon density at 
low temperatures and $\mu_B \gtrsim 3 m_\pi/2$, where $m_\pi$ is 
the Goldstone pion mass. We elucidate the nature of the non-analyticity 
behind this sharp rise in the density by a comparison of reweighting 
results with a Taylor expansion of high order. While at first sight this 
non-analytic behavior becomes apparent at the same position where the pion 
condensation transition takes place in the phase-quenched theory, but the 
nature of the non-analyticity in the two theories appears to be quite different: 
While at nonzero isospin density the data are consistent with a 
genuine thermodynamic (branch-point) singularity, 
the results at nonzero
baryon density point to an essential singularity 
at $\mu_B=0$. The effect is absent for four flavors of degenerate quarks, 
where rooting is not used. For the two-flavor case, we 
show numerical evidence that the magnitude of the effect 
diminishes on finer lattices. We discuss the implications of this
technical complication on future studies of the QCD phase diagram.
\end{abstract}

\maketitle

\section{\label{sec:intro}Introduction}
Despite decades of effort, the determination of the phase diagram of QCD 
on the temperature ($T$) -- baryochemical potential ($\mu_B$) plane
remains an unsolved problem, due to the complex action problem hampering 
first-principle lattice QCD calculations. Nevertheless, several 
workarounds have been
proposed and utilized to 
obtain information at nonzero $\mu_B$, such 
as a Taylor expansion around $\mu_B=0$~\cite{Gavai:2003mf,Allton:2005gk,
  MILC:2008reg,Borsanyi:2011sw,Borsanyi:2012cr,
  Bellwied:2015lba,Ding:2015fca,Bazavov:2017dus,HotQCD:2018pds,
  Giordano:2019slo,Bazavov:2020bjn}, analytic continuation from imaginary
chemical potentials~\cite{deForcrand:2002hgr,DElia:2002tig,DElia:2009pdy,
  Cea:2014xva,Bonati:2014kpa,Cea:2015cya,Bonati:2015bha,
  Bellwied:2015rza,DElia:2016jqh,Gunther:2016vcp,Alba:2017mqu,
  Vovchenko:2017xad,Bonati:2018nut,Borsanyi:2018grb,Bellwied:2019pxh,
  Borsanyi:2020fev}, and different reweighting techniques~\cite{
Barbour:1997ej,Fodor:2001au,Fodor:2001pe,Fodor:2004nz,deForcrand:2002pa,Alexandru:2005ix,Giordano:2020roi,Borsanyi:2021hbk, Fodor:2007vv,Endrodi:2018zda,Borsanyi:2021hbk,Borsanyi:2022soo}. 

Since all of these workarounds require very large statistics, rooted 
staggered fermions~\cite{Sharpe:2006re,Kronfeld:2007ek} are the most popular 
fermion discretization in the literature for being the most 
computationally efficient. In fact, 
we are not aware of any results with physical quark masses on finite baryon density QCD 
from non-staggered formulations. Similarly, continuum extrapolation at finite baryon 
density has only ever been attempted with rooted staggered fermions. 
It is unfortunate then, that the theoretical justification of the application of rooted 
staggered fermions at finite chemical potential is not fully 
settled~\cite{Golterman:2006rw,Giordano:2019gev}. 

For simplicity, we discuss the case when a chemical potential 
is only introduced for degenerate light quarks, and not for the
strange quark. In this case one has to deal with complex square 
roots instead of fourth roots. Since the staggered determinant is complex 
at real chemical potential, one must find a way to resolve the sign 
ambiguity in the complex square root function. One 
possible way is to demand the determinant to be a continuous function 
of the chemical potential along the real axis~\cite{Fodor:2004nz}. 
In Ref.~\cite{Golterman:2006rw} a simple counting
argument was given, that suggests that this procedure leads to 
cut-off effects of the order $\mathcal{O} \left( a \right)$, where
$a$ is the lattice spacing. The counting argument is 
based on the observation that the taste multiplets in the Dirac 
spectrum are of size $\mathcal{O}(a)$. Unfortunately, it gives 
no indications about the chemical potential dependence of 
these cut-off effects.

\begin{figure}
\centering
\includegraphics[width=0.95\linewidth]{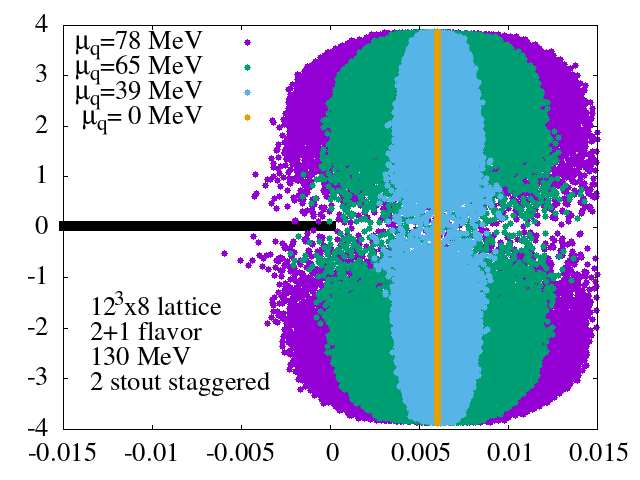}
\vspace{-0.2cm}
\caption{
Illustration of the Dirac spectrum at finite baryo-chemical potential.
The spectrum, obtained with dense linear algebra, is based on one dynamical
staggered lattice configuration of size $12^3 \times 8$ with $m_\pi=135~\textrm{MeV}$, using the
2stout action. The thick line shows the standard choice for the
branch cut of the complex square root. 
\label{fig:diracspectrum}
}
\end{figure}

Since the branch point singularities of the complex 
square root function are located at the zeros of its argument, 
the problems caused by the ambiguity in rooting are expected to be severe at values of the chemical
potential near a complex zero of the (unrooted) staggered determinant.
At zero temperature, this is 
expected to happen for light-quark chemical potentials
$\mu_q=\frac{\mu_B}{3} \gtrsim \frac{m_\pi}{2}$~\cite{Gibbs:1986hi,Fodor:2007ga}. 
See Fig.~\ref{fig:diracspectrum} for an illustration
at a temperature in the confined phase.
How thermodynamic observables are affected by these 
branch point singularities has, however, never been 
studied in detail. This is the purpose of this work.

The structure of this paper is the following: First, in 
section \ref{sec:reweighting} we revisit 
reweighting methods, including methods that are free from 
an overlap problem~\cite{Giordano:2020roi,Borsanyi:2021hbk,Borsanyi:2022soo}. 
The use of such methods is important, as they allow us to 
safely rule out an
overlap problem in our results.

In section \ref{sec:rise}, we show that rooted staggered fermions 
predict a sharp rise in the 
light-quark density around $\mu = \mu_B/3 \gtrsim  m_\pi/2$
if we resolve the rooting ambiguity by requiring that the 
determinant is a continuous
function at real $\mu_B \geq 0$.
This sharp rise occurs
at values of the quark chemical potential where, in the phase-quenched theory, 
the isospin density shows a sharp rise~\cite{Brandt:2018omg}. In the latter case, which is equivalent to a theory with nonzero
isospin chemical potential, the rise of the isospin
density signals the condensation of pions~\cite{Son:2000xc,Brandt:2017oyy}.

At first sight, it thus appears that the baryon chemical potential 
ensemble has some remnant of the pion condensation transition at nonzero 
isospin chemical potential. 
In fact, this behavior was seen before in Ref.~\cite{Brandt:2019ttv}, where 
the authors attempted to reweight ensembles 
generated at a nonzero isospin chemical 
potential to a nonzero baryochemical potential. 
At that point 
in time, however, it could not be determined whether the cause of this issue 
was an overlap problem from reweighting or a problem with staggered rooting. With the methods discussed in section~\ref{sec:reweighting},
we can now safely rule out an overlap problem.

To confirm that the cause of the issue is indeed rooting, we show 
results from simulations with four flavors of degenerate quarks 
in section \ref{sec:4flavor}.
When the chemical potential is nonzero for only two of the 
four flavors (which requires rooting), the rapid rise of the density at
half the pion mass is again present. On the other hand, 
if the chemical potential is the same for all four flavors (which 
requires no rooting), the issue is not observed, as we cannot
observe such a sharp rise in the 
density.

The question then remains: 
To what extent can the observed increase in the density at 
nonzero $\mu_B$ 
be considered a remnant of the pion condensation transition?
To make the question more concrete: is the analytic structure of the free energy at nonzero 
$\mu_B$ similar
to what one expects for a thermodynamic transition, or
is there any difference?
In section \ref{sec:analytic}, we elucidate the nature of the non-analyticity 
introduced by staggered rooting 
by comparing
results from reweighting methods to results obtained by performing a Taylor expansion of the partition function at zero chemical potential, which we calculated to unprecedentedly 
high orders. At low temperature, in spite of the Taylor coefficients computed 
at zero chemical potential being exactly the same as the Taylor coefficients 
of the expansion of the reweighted pressure, the two methods appear to converge to 
different curves. This suggests that the rooted staggered free energy is non-analytic in the chemical potential, with a non-analytic term
having an identically zero Taylor expansion. Such a behavior is not expected
around zero on thermodynamical grounds, since the QCD transition for vanishing
baryon density is a crossover.
This behavior is also very different from the non-analytic behavior at a 
typical thermodynamic transition or crossover: Even though these are also characterized by a sudden rise in the
density as a function of the chemical potential, the behavior of the Taylor series is very different. For a phase transition or
a crossover, the Taylor expansion should diverge near the transition point~\footnote{The Taylor expansion should also 
diverge in the vicinity of a crossover, due to nearby Lee-Yang zeros in the complex 
chemical potential plane.}. 
We also demonstrate this contrasting behavior using lattice data by 
comparing a high order Taylor expansion with reweighting at 
a nonzero isospin density.

Next, in section \ref{sec:continuum}, we numerically extract the 
non-analytic part (under some assumptions) for a discretization with strongly
suppressed taste breaking - known as 4HEX improved staggered fermions~\cite{Borsanyi:2020mff} - for three different lattice spacings ($N_\tau=6,8,10$)
and observe a strong decrease in the magnitude of the non-analytic 
part, which is an indication that this sharp rise in the 
density for $\mu_B\gtrsim 3 m_\pi/2$ is a cut-off effect.

In the final section we summarize our results and discuss several strategies which 
future simulation projects can use to overcome this roadblock introduced by staggered 
rooting. 

\section{\label{sec:reweighting}Rooted staggered quarks at finite $\mu_B$} 
The dimensionless pressure is related to the grand canonical partition function $\mathcal{Z}$ of a thermodynamic system via:
\begin{equation}
\label{eq:p}
    \hat{p} \equiv \frac{p}{T^4} = \frac{1}{T^3 V} \log \mathcal{Z}\rm,
\end{equation}
where $T$ is the temperature and $V$ is the spatial volume.
In this paper we deal with the grand canonical partition functions of QCD and related models at finite quark chemical potential, studying in particular the chemical potential dependence of the quark density-to-chemical potential ratio, defined as:
\begin{equation}
\label{eq:nL}
    \frac{\hat{n}_L}{\hat{\mu}_q} \equiv \frac{1}{\hat{\mu}_q} \frac{\partial \hat{p}}{\partial \hat{\mu}_q} = \frac{1}{\mu_q T V}\frac{\partial \log \mathcal{Z}}{\partial \mu_q}\rm,
\end{equation}
where $\hat{n}_L = n_L/T^3$ and $\hat{\mu}_q=\mu_q/T$. We will apply 
equations~\eqref{eq:p} and ~\eqref{eq:nL} for different theories, with different 
flavor content and choices for the chemical potentials throughout this manuscript.

For the case of $N_f=2+1$ flavors of rooted staggered fermions, the grand canonical partition function reads schematically:
\begin{widetext}
\begin{equation}
\label{eq:staggeredZ}
\mathcal{Z}_{2+1}(T,\mu_q) = \int \mathcal{D}U \det M^{1/2}(U,m_u,\mu_q) \det M^{1/4}(U,m_s,0) e^{-S_{\textrm{YM}}(U)} \rm.
\end{equation}
\end{widetext}
Here $M$ is the massive staggered operator,  $m_u$ and $m_s$ are the light- and strange quark masses, respectively, $\mu_q$ is the chemical potential of the light quarks, while we set the strange quark chemical potential $\mu_s=0$, which is the setup we use throughout the 
paper. Moreover, $S_{YM}$ is the discretized Yang-Mills action and $U$ are the link variables.

Since the integrand in Eq.\eqref{eq:staggeredZ} is complex, the partition function at finite chemical potential cannot be numerically simulated using standard importance-sampling methods. In this paper we side-step this problem using reweighting techniques, i.e., performing a numerical simulation using importance sampling of a related theory free from the complex-action problem, and then suitably rescaling the weight of each configuration so that it matches the (complex) one found in the target theory.
We will obtain the same result with three different reweighting schemes, which we describe in the following:

\paragraph{Reweighting from $\mu_B=0$:} In what is arguably the simplest reweighting technique, one generates 
configurations at $\mu_q=0$ and calculates the partition function, its logarithm and the derivatives of its logarithm 
by starting with the formula:
\begin{equation}
    \label{eq:Glasgow}
\frac{\mathcal{Z}_{2+1} (\mu_q)}{\mathcal{Z}_{2+1} (\mu_q=0)} 
= \left\langle \frac{\det M^{1/2}(U,m_u,\mu_q)}{\det M^{1/2} (U,m_u,0)} \right\rangle_{\mu_q=0} \rm{,}
\end{equation}
where $\left\langle \dots \right\rangle_{\mu_q=0}$ denotes expectation values calculated at $\mu_q=0$. Derivatives of the pressure follow by simply differentiating the natural logarithm of Eq.~\eqref{eq:Glasgow}.

\begin{figure}[t]
  \begin{center}
          \includegraphics[angle=270,width=0.95\linewidth]{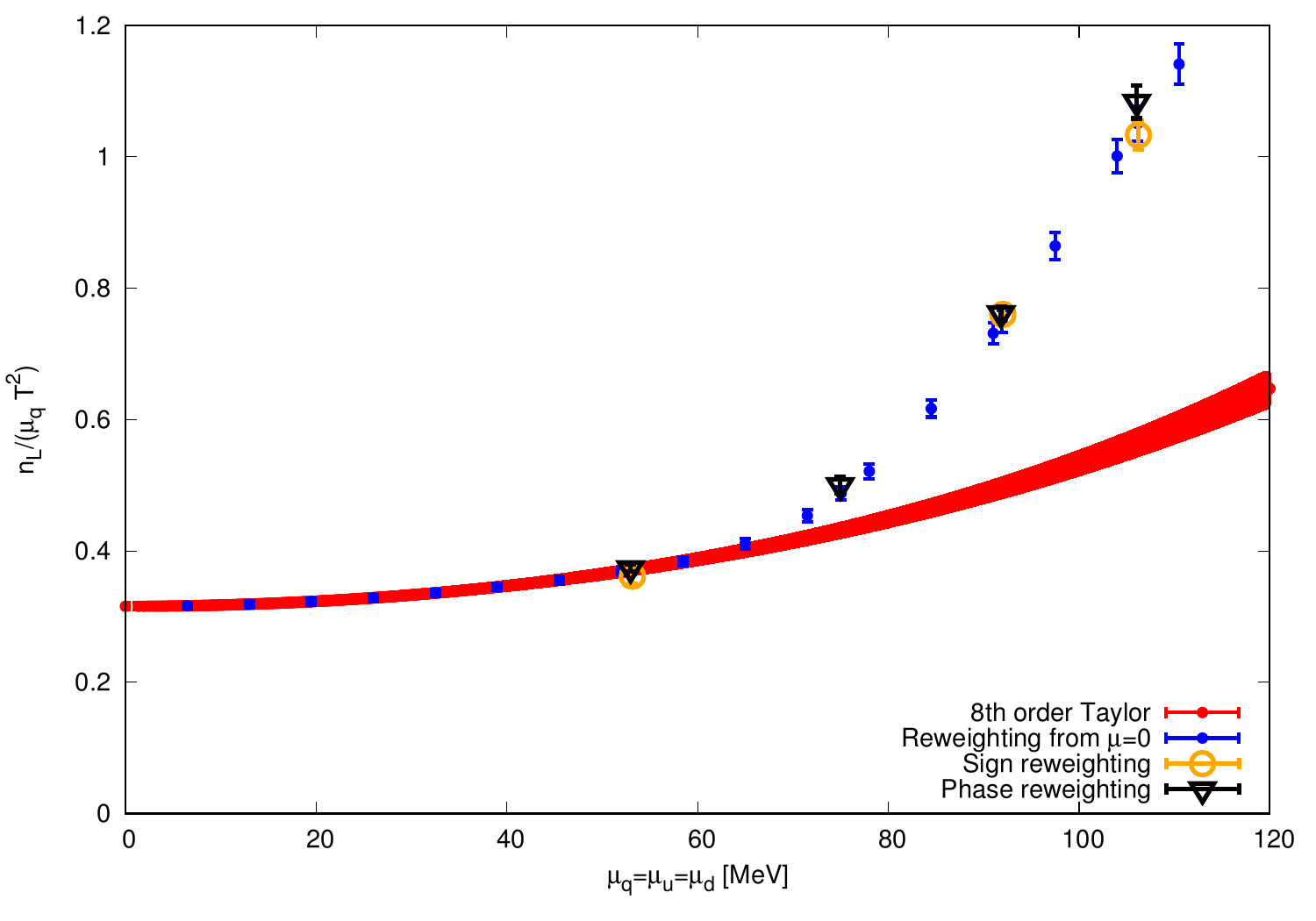}
          \vspace{-0.2cm}
  \end{center} 
  \caption{
      A rapid rise in the light-quark density-to-chemical potential 
      ratio as a function of $\mu_q$ around $\mu_q=m_\pi/2$ with rooted staggered fermions. We used the 2stout improved action, with a lattice
      size of $16^3 \times 8$ and a temperature of $T=130$MeV. To 
      lead the eye and see where
      the increase sets in, an 8th-order Taylor expansion is also shown. 
      Reweighting from $\mu_q=0$, phase reweighting, and sign reweighting (see section III) all give the same result, indicating a lack of an overlap problem.  
      \label{fig:fake_transition}
  }
\end{figure}

The ratio of determinants in Eq.~\eqref{eq:Glasgow}  can be conveniently calculated using the reduced matrix formalism~\cite{Hasenfratz:1991ax}. The staggered reduced matrix is a complex
matrix of size $6N_s^3\times 6N_s^3$, where $N_s^3$ is the spatial lattice volume in lattice
units. The reduced matrix is a function
of the gauge links as well as the quark mass, but, most importantly, it is
independent of the chemical potential. Using the notations of Ref.~\cite{Giordano:2019gev} one can
express the staggered determinant of the four-flavor theory for a given configuration ($U$) at any quark
chemical potential ($\mu_q$) using only the eigenvalues ($\xi_i$) of the reduced matrix as
\begin{equation}
\label{eq:reduced_matrix}
\frac{\det M(U,m,\mu_q)}{\det M(U,m,0)}=e^{-3N_s^3\mu_q/T} \prod_{i=1}^{6N_s^3} \frac{\xi_i[m,U] - e^{\mu_q/T}}{\xi_i[m,U] -1}\,.
\end{equation}
The eigenvalues ($\xi_i$) can be computed using dense linear algebra 
packages \cite{icl:443}.

The rooting procedure for a complex fermion determinant is inherently ambiguous, and 
the ratio of rooted determinants in Eq.~\eqref{eq:Glasgow} is not well defined until this ambiguity is resolved. A reasonable choice is to compute this ratio by taking the square root of Eq.~\eqref{eq:reduced_matrix} eigenvalue by eigenvalue~\cite{Fodor:2001pe,Giordano:2019gev}:
\begin{equation}
\frac{\det M^{1/2}(U,m,\mu_q)}{\det M^{1/2}(U,m,0)}:=e^{-3N_s^3\mu_q/T} \prod_{i=1}^{6N_s^3} \sqrt{\frac{\xi_i[m,U] - e^{\mu_q/T}}{\xi_i[m,U] -1}} \rm.
\label{eq:rooted_reduced}
\end{equation}
The branch cut of the complex square root on the right hand side is put on the negative real axis.
Notice that none of the fractions under the square roots will ever cross
the branch cut of the square root function as long as $\mu_q$ is real. 
As a consequence, the rooted determinant as defined in Eq. ~\eqref{eq:rooted_reduced}
continuously connects to
the positive real root of the determinant at $\mu_q=0$ starting from any 
real value of $\mu_q=\mu_B/3$.

While Eq.~\eqref{eq:Glasgow} is exact for infinite statistics, 
the probability distribution of the 
weights $\frac{\det^{1/2} M(U,m_u,\mu_q)}{\det^{1/2} M(U,m_u,0)}$ can be 
heavy tailed, and thus hard to sample (an overlap problem). 
It is therefore hard to judge the reliability of this reweighting approach on its own, without also using other techniques.
To cross-check the reweighting method from $\mu_q = 0$ we utilize two other reweighting schemes for which the weights take values from a compact domain.  By construction, probability distributions with a compact support have no tails, and therefore no overlap problem.

\paragraph{Phase reweighting:} For this reweighting scheme, the simulated ensemble 
is the phase-quenched ensemble, with quark determinant replaced by its absolute 
value. The partition function of the phase-quenched ensemble reads:
\begin{widetext}
\begin{equation}
\label{eq:ZPQ}
    \mathcal{Z}_{2+1}^{PQ}(T,\mu_q) = \int \mathcal{D}U | \det M^{1/2}(U,m_u,\mu_q) | \det M^{1/4}(U,m_s,0) e^{-S_{\textrm{YM}}(U)} \rm{,}
\end{equation}
\end{widetext}
and is identical to the partition function with a finite isospin chemical 
potential, with $\mu_q=\mu_u=-\mu_d$.
The partition function at finite baryochemical potential is obtained 
from the phase-quenched ensemble as 
\begin{equation}
    \frac{\mathcal{Z}_{2+1}}{\mathcal{Z}_{2+1}^{PQ}} = \left\langle \frac{\det M^{1/2}(U,m_u,\mu_q)} {| \det M^{1/2}(U,m_u,\mu_q) |} \right\rangle_{PQ} \rm{,}
\end{equation}
where $\left\langle \dots \right\rangle_{PQ}$ denotes the expectation value in 
the phase-quenched theory, defined by Eq.~\eqref{eq:ZPQ}.
The weights in this case are pure phases, i.e., the reweighting factors are 
elements of the unit circle, which form a compact 
domain, on which no long-tailed distributions are possible.

\begin{figure*}[t!]
  \centering
          \includegraphics[width=0.49\linewidth]{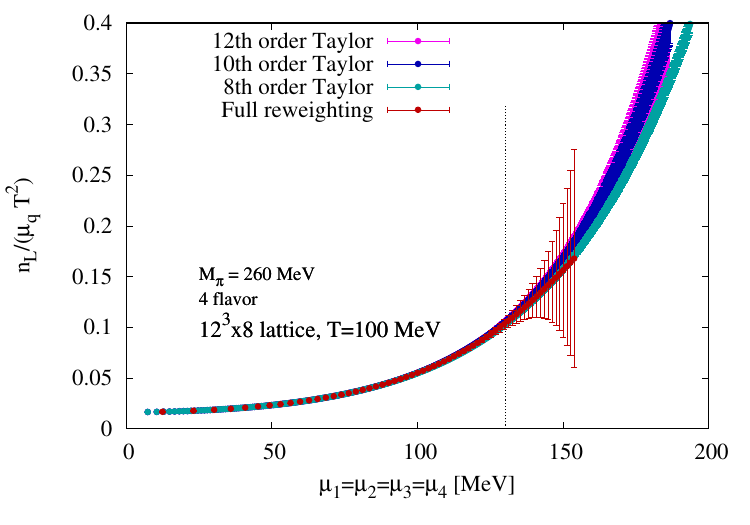}
          \includegraphics[width=0.49\linewidth]{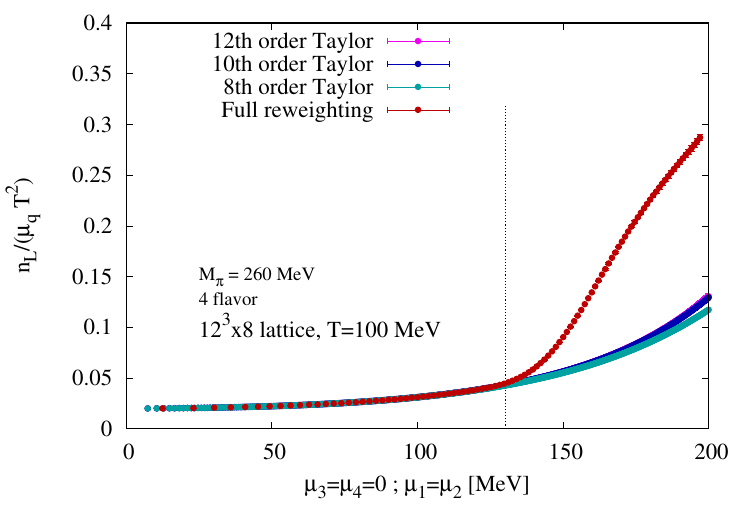}
  \caption{
      The density to chemical potential ratio for four flavors of quarks. Left: all quarks having the same chemical potential (this choice requires no rooting)
      Right: the chemical potential is introduced only for two of the four flavors, while the other two flavors remain at zero chemical potential (this choice requires rooting).
      Both cases where calculated using the same $\mu_q=0$ ensemble. The vertical line corresponds to $\mu_q = m_\pi/2$.
      \label{fig:Nf4}
  }
\end{figure*}

While the idea of phase reweighting has been around for 
decades, the first simulations in the actual phase-quenched 
ensemble have been carried out only two 
years ago~\cite{Borsanyi:2021hbk}. Previous studies always included 
an explicit symmetry-breaking term~\cite{Kogut:2002zg}, usually denoted $\lambda$, which 
amounts to the substitution $|\det M|^{1/2} \to \det (M^\dagger M + \lambda^2)^{1/4}$ in
Eq.~\eqref{eq:ZPQ}. In the $\lambda\to0$ limit this formulation allows
theoretically clean investigations of pion condensation, making it a preferable choice
for nonzero isospin density. However, when simulations are performed with the intent to
subsequently reweight to nonzero $\mu_B$,
the $\lambda$ term is not desirable, as it leads to an 
overlap 
problem in the $\lambda \to 0$ reweighting step.
How simulations at $\lambda=0$ are performed is explained in Ref.~\cite{Borsanyi:2021hbk}.

\paragraph{Sign reweighting:} Here the partition function is defined as:
\begin{widetext}
\begin{equation}
\label{eq:ZSQ}
    \mathcal{Z}_{2+1}^{SQ}(T,\mu_q) = \int \mathcal{D}U | \operatorname{Re} \det M^{1/2}(U,m_u,\mu_q) | \det M^{1/4}(U,m_s,0) e^{-S_{\textrm{YM}}(U)} \rm{,}
\end{equation}
\end{widetext}
and the ratio of the desired finite baryon density vs the sign-quenched partition function is:
\begin{equation}
    \frac{\mathcal{Z}_{2+1}}{\mathcal{Z}_{2+1}^{SQ}} = \left\langle \frac{\operatorname{Re} \det M^{1/2}(U,m_u,\mu_q)} {| \operatorname{Re} \det M^{1/2}(U,m_u,\mu_q) |} \right\rangle_{SQ} \rm{,}
\end{equation}
where $\left\langle \dots \right\rangle_{SQ}$ denotes the expectation value in 
the sign-quenched theory, defined by Eq.~\eqref{eq:ZSQ}.
The reweighting factors can only take two values: $+1$ or $-1$. Hence the 
distribution of the weights is a one parameter
probability distribution (the Bernoulli distribution). Again, by construction, there 
are no tails and thus no overlap problem. Note that the substitution of the 
quark determinant to its real part is not allowed in generic expectation values,
as the path integral representation of expectation value will in general involve 
the determinant itself. The substitution is allowed, however, 
for the class of observables we consider here: i.e., observables that can be defined 
as real derivatives of the partition function with respect to real parameters, like 
the quark mass or the chemical potential.

\section{\label{sec:rise}A rise in the density at $\mu_B \gtrsim 3 m_\pi/2$} 

We have recently shown that for the equation of state 
of the quark gluon plasma in the range $\mu_B/T \leq 3$, reweighting gives 
compatible results with analytic continuation from purely imaginary chemical 
potentials~\cite{Borsanyi:2022soo}. Thus, in the currently experimentally accessible 
range (the range of the RHIC Beam Energy Scan, phase two) the equation of state of 
the quark-gluon plasma is under control.
In particular, the resummations introduced in Refs.~\cite{Borsanyi:2021sxv,Borsanyi:2022qlh}, an 8th-order Taylor expansion and the overlap 
problem free reweighting techniques of Refs.~\cite{Giordano:2020roi,Borsanyi:2021hbk} 
agree in this range for $\frac{\hat{n}_L}{\hat{\mu}_q}$.
The lowest temperature studied in 
Ref.~\cite{Borsanyi:2022soo} was 145~MeV.

When attempting to extend these reweighting studies to lower temperatures, we have 
noticed a sharp rise in the density as a function of the chemical potential
at around $\mu_B = 3m_\pi/2$. This is shown for a temperature of $T=130$~MeV in Fig.~\ref{fig:fake_transition}. Results in this plot were obtained with a lattice size of $16^3 \times 8$. We use 
physical quark masses, using a tree-level improved gauge action and 2 steps of stout smearing~\cite{Morningstar:2003gk} with smearing parameter $\rho=0.15$ applied 
to the links entering the staggered Dirac operator, which  we will call the 2stout action~\cite{Aoki:2006we, Borsanyi:2010bp} from now on.
To better see the onset of this sharp increase, we also show the 8th-order 
Taylor expansion, as a smooth baseline. 
Note that in this work we compute Taylor expansions around $\mu_q=0$ using the reduced matrix formalism~\cite{Hasenfratz:1991ax}, without employing stochastic estimators.

This sharp rise in the density is very similar to the way the isospin density behaves 
at the pion condensation transition~\cite{Brandt:2018omg}, a transition 
that happens in the phase-quenched theory 
around the same value $\mu_q=m_\pi/2$ of the quark chemical potential. 
In fact, it was seen before in the literature that at $\mu_B=3m_\pi/2$, something
that looks similar to a phase transition takes place. This was observed when 
attempting to 
reweight to a nonzero baryochemical potential from a 
nonzero isospin chemical potential in Ref.~\cite{Brandt:2019ttv}. 
The main difference between that work and ours
is that we did not introduce an infrared regulator 
that lifts the masse of the 
Goldstone boson
that emerges as a result of the symmetry-breaking transition. We avoided using
this regulator to guard our simulations 
from an additional overlap problem.  

The strong deviation of the reweighted results from the Taylor expansion in Fig.~\ref{fig:fake_transition}
prompts for cross-checks. Indeed, we computed the sharp rising baryon density using the three different
reweighting schemes discussed in the previous section, all giving the same result, and ruling out an overlap problem. 
Since we have established the reliability of 
these schemes, we will exclusively use reweighting 
from $\mu_q=0$ in 
the further sections of the paper, to keep the
computer time budget manageable.

\section{\label{sec:4flavor}The four-flavor theory} 

If the observed sharp rise in the quark density is due to staggered rooting, it should 
be absent in the four-flavor theory, where rooting is
absent and the partition function with staggered fermions is given by
\begin{equation}
    \mathcal{Z}_{4}(T,\mu_q) = \int \mathcal{D}U \det M(U,m,\mu_q) e^{-S_{\textrm{YM}}(U)} \rm{,}
\end{equation}
where $m$ is the quark mass for all four flavors. We can calculate expectation values of
observables in this theory 
by reweighting from zero chemical potential, similarly to the 2+1 flavor case.

On the other hand, the sharp rise in the density should be observed if we only introduce a chemical potential for only two of the four flavors:
\begin{widetext}
\begin{equation}
    \mathcal{Z}_{2+2}(T,\mu_q) = \int \mathcal{D}U \det M^{1/2}(U,m,\mu_q) \det M^{1/2}(U,m,0) e^{-S_{\textrm{YM}}(U)} \rm{.}
\end{equation}
\end{widetext}

We have performed simulation at $\mu_q=0$ for the four-flavor theory, with a quark masses corresponding 
to a pion mass of $m_\pi=260$~MeV~\footnote{For the four-flavor theory, it is hard to set the pion mass smaller with staggered fermions, due to the bulk phase discussed in Refs.~\cite{Lee:1999zxa,Aubin:2004dm,Cheng:2011ic,Aubin:2015dgk,Kotov:2021mgp}}, a temperature $T=100$~MeV 
and a lattice volume of $12^3 \times 8$.  
The scale was set using the $w_0$ scale of Ref.~\cite{BMW:2012hcm}.
We calculated the eigenvalues of the reduced matrix
for 0.9 million configurations.
With this pion mass the inflection point of the renormalized chiral condensate gives a cross-over temperature of approximately 135~MeV.
We then proceeded to calculate the 
ratio $\hat{n}_L/\hat{\mu_q}$ for both assignments of the chemical potentials, i.e., when all four quarks get the same chemical 
potential vs when only two of them do. The 
results can be seen in Fig.~\ref{fig:Nf4}. 
While the four-flavor case $\mu_1=\mu_2=\mu_3=\mu_4=\mu_q$ becomes quite noisy
at large $\mu_q$, the 
difference between the two cases is rather clear to see. 
In the four-flavor case the Taylor expansion and the full reweighting match 
within errors.
In the case of $\mu_1=\mu_2=\mu_q$ with $\mu_3=\mu_4=0$, the reweighted curve rises 
sharply near $\mu_q = m_\pi/2$, and there is a very clear
deviation between the Taylor and reweighted curves. This resembles the sharp
rise in the quark density that we observed in the $2+1$-flavor case. 
The statistical errors for the (unrooted) four-flavor case are admittedly large at large $\mu_q$. The error grows large precisely at the chemical potential where 
the rooted case takes a sharp turn, indicating large cancellations. Note that
the four-flavor theory is expected to be noisier then the two-flavour theory, 
since the exponential severity of
the sign problem has a factor of $N_f^2$ in the exponent~\cite{Borsanyi:2021hbk}, i.e. $\log \frac{\mathcal{Z}_{N_f}}{\mathcal{Z}_{N_f}^{PQ}} \propto N_f^2$.

These findings are in line with our assumption that 
the sharp rise in the light-quark density 
for $\mu_B \gtrsim 3m_\pi/2$ is caused by 
staggered rooting.


\section{\label{sec:analytic}Analytic structure and pion mass dependence} 

\begin{figure*}[t]
  \centering
          \includegraphics[width=0.49\linewidth]{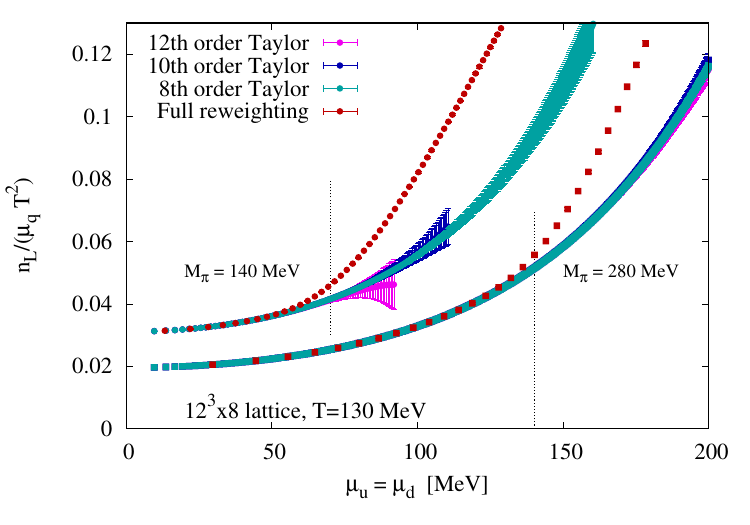}
          \includegraphics[width=0.49\linewidth]{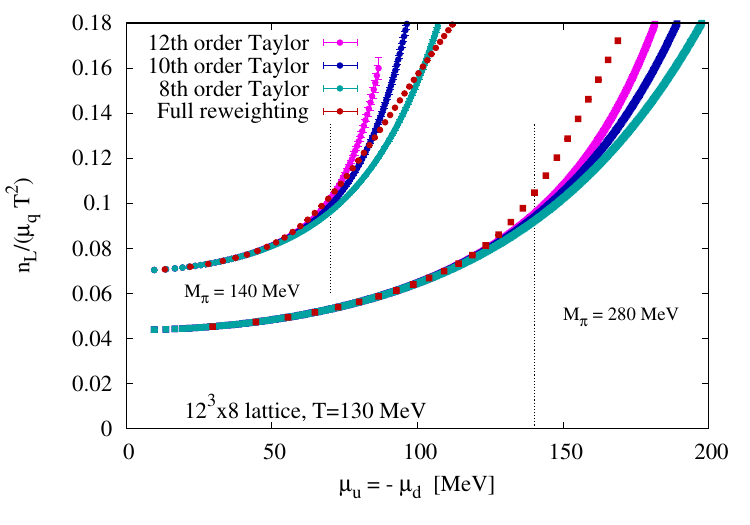}
  \caption{
      The full reweighted results for the light-wark density compared with several high orders of the Taylor expansion for the case of a baryon (left panel) and isospin (right panel) chemical
      potential for two different values of the pion mass. 
      While the density itself looks very similar in the two cases, with 
      a rapid rise for $\mu_q \gtrsim m_\pi/2$, the two singularities
      appear to be very different: At a finite isospin chemical potential - where the singularity of the free energy is due to a true phase transition - 
      the Taylor expansion converges to the full 
      reweighting result below $\mu_q \approx m_\pi/2$, and diverges above. On the other hand, at a finite baryon chemical potential, the Taylor expansion appears to
      converge, but above $\mu_q \approx m_\pi/2$ it converges to a different curve than the full reweighting one. 
      This hints at a very different analytic structure.
      \label{fig:analytic_structure}
  }
\end{figure*}

Next, we try to answer the question: to what extent is this sharp rise in the
quark density in the nonzero baryo-chemical potential ensemble similar to or
different from the sharp rise in the isospin density in the nonzero isospin
ensemble. More precisely: Is the complex singularity behind these two behaviors 
of the same type?

In order to clarify the analytic structure behind the observed 
rise in the density, we have performed high statistics simulations
on $12^3 \times 8$ lattices with the 2stout action at physical quark masses, 
and measured Taylor coefficients of the pressure to 12th order in the chemical potential 
both for physical
quark masses and also for four times the physical value of the quark 
masses (which correspond to twice the 
physical pion mass to leading order in chiral perturbation theory). When converting
results to MeV in the heavy-pion case, we assumed that the lattice scale does not 
change when changing the light quark mass. This assumption is approximately
correct as the scale depends more strongly on the mass of the strange quark.
Our results can be seen in the left panel of Fig.~\ref{fig:analytic_structure}. 

For comparison, we 
also show the Taylor expansion and the full reweighting for the case of an isospin
chemical potential (with no explicit symmetry breaking parameter) in the right panel
of Fig.~\ref{fig:analytic_structure}. For the case of an isospin chemical potential 
the rapid rise in the density is due to a second order phase 
transition~\cite{Son:2000xc,Brandt:2017oyy} at $\mu_q \approx m_\pi/2$. In such a case
the free energy is analytic at $\mu_q=0$ and has a branch-point 
singularity at the nearest Lee-Yang 
zero~\cite{Lee:1952ig,Deger:2019mgo,Giordano:2019slo,Brange:2023cdf}, limiting 
the radius of convergence. At large enough orders in the finite-volume 
Taylor expansion a crossover, a second order transition and a first order
transition are not that different, as the position of a Lee-Yang zero must have 
a nonzero imaginary part $\Delta_{LY}$, which for a true phase transition vanishes
in the thermodynamic limit, while for a crossover remains finite. In this situation
the radius of convergence in a finite volume is given approximately
by $R \approx \sqrt{\mu_c^2 + \Delta_{LY}^2}$, where $\mu_c$ is the 
chemical potential of the transition in the infinite-volume limit. 
For large enough orders, the nonzero $\Delta_{LY}$ leads to a complicated 
sign structure for the high order coefficients~\cite{Giordano:2019slo}. 
We do not observe this up to
12th order in the Taylor expansion, however. Rather, at finite isospin chemical potential
all Taylor coefficients up to this order are positive, and the radius of
convergence can be estimated with a simple ratio estimator to be close to 
the expected $R \approx m_\pi/2$. Thus the 
Taylor expansion should converge to the full result until the transition point, 
above which it should diverge. This is exactly what is seen in the right panel
of Fig.~\ref{fig:analytic_structure}, at both of the simulated pion masses.
The one important difference is that for the larger pion mass, the Taylor 
expansion seems to converge more slowly. This is not surprising, as the expansion 
parameter is $\mu_q/T$, which is larger at the transition point for a heavier pion.

Such divergent behavior is typical for phase transitions. It is very different
from the behavior we observe at nonzero baryon density, seen in the left panel of Fig.~\ref{fig:analytic_structure}.
Here, the Taylor expansion appears to converge, but it converges to a different curve
from the one obtained with full reweighting. Let us emphasize that the Taylor coefficients and the 
reweighting curve were obtained using the same gauge ensembles, and so the 
Taylor coefficients are exactly the Taylor coefficients of the reweighted curve
at $\mu_q=0$.
Also note that while the point of divergence between the reweighting and Taylor curves does seem
to scale with the pion mass, the two curves actually start to diverge already somewhat below 
$\mu_q=m_\pi/2$. This points to quite different non-analytic behavior, compared to the case 
of a nonzero isospin density. One example of a non-analytic term in the 
free energy which could produce such a
behavior is 
\begin{equation}
\label{eq:essential_singularity}
A \mu_q^{\alpha} e^{-\frac{\Lambda^2}{\mu_q^2}}\rm{,}
\end{equation}
where $A$, $\alpha$ and $\Lambda$ are parameters. 
A term of this form in the free energy would not affect the Taylor expansion at all, 
but it would lead to a sharp increase in the density at some value of $\mu\approx \Lambda$.  If $\Lambda$ also scaled with the pion mass, then so would 
the value at which the sharp increase appears.
If such terms are indeed the cause of the sharp rise of the light-quark density, 
then the non-analyticity of the free energy with rooted staggered fermions 
would actually be located at $\mu_q=0$ and not near $\mu_q=m_\pi/2$. 

While it is impossible to prove that the free energy has a term of this form by using only numerical simulations, there are two pieces of evidence that support this conjecture.

First, one can easily construct two particular schemes of the complex rooting that differ in the free energy in the conjectured form.\\
i) Consider Eq.~(\ref{eq:rooted_reduced}), which is designed to be continuous 
in real $\mu_q$ and
has been used, e.g., in Fig.~\ref{fig:analytic_structure} for the ``full reweighting'' data. \\
ii) Consider the square root of the determinant where we always take the root with a positive real part.\\

Notice that the second definition will never lead to a sign problem.
In fact, the second definition gives the sign-quenched
partition function (see Eq.~(\ref{eq:ZSQ})) of the first definition.
One can
show (see the Appendix of Ref.~\cite{Borsanyi:2021hbk}) that the 
difference between the two free energies is of the conjectured form, with $\alpha=3$,
\begin{equation}
f - f_{SQ} \sim A \mu_q^3 e^{-B/\mu_q^2}\rm{.}
\end{equation}
Thus, the Taylor series of i) and ii) schemes are identical and the two free-energies
near $\mu_q=0$ differ in a function with an essential singularity. 
One therefore expects that
different choices of the rooting procedure lead to different singularities. 
However, we are not aware of any choice where analyticity at $\mu_q=0$ is guaranteed.

Second, the conjectured form of the non-analytic part fits the lattice 
simulation data well. 
Thus, while the exact functional 
form is hard to determine from numerical data alone, the numerical evidence 
points to the rooted staggered free energy having an essential singularity 
at $\mu_q=0$.

Finally, we point out that an essential singularity of the form
shown in eq.~\ref{eq:essential_singularity} is (strictly 
speaking) only possible for infinite statistics. For any 
finite statistics, the leading singularity in the free
energy should be the branch-point singularity given by
the closest of the determinant zeros (complex logarithms of the
$\xi_i$ eigenvalues of the reduced matrix in eq.~\eqref{eq:reduced_matrix}) 
in the ensemble. In the limit of infinite statistics the 
branch-points could get arbitrarily close to $\mu_q=0$.
Then the accumulation of these branch-point singularities 
could produce an essential singularity. 
Strictly speaking this means that at finite statistics,
the Taylor series has a finite radius of convergence, given by the 
closest branch-point singularity, while in the limit of infinite 
statistics, the radius of convergence tends to zero, in the sense of an 
essential singularity at $\mu_q=0$. Only in the limit of infinite
statistics, is the Taylor expansion oblivious to the branch points.
We have seen such behavior 
while accumulating the statistics for Fig.~\ref{fig:analytic_structure}. 
When we had only around 100k configurations, the 
Taylor coefficients $\chi^L_{10}$ and $\chi^L_{12}$ 
looked like they had a non-zero value within errors, 
and estimators of the radius of convergence gave 
results between $60$ and $70$MeV. After doubling the
statistics, the signal for $\chi^L_{10}$ and $\chi^L_{12}$
disappeared, and behavior similar to the one shown in the 
left panel of 
Fig.~\ref{fig:analytic_structure} emerged, which then
remained stable after doubling the statistics two more times.
Thus, the estimators of the radius of convergence with the
smaller statistics were most likely dominated by a 
few configurations with
close-by branch point singularities, and only after 
substantially increasing the statistics did the behavior
become apparently consistent with an essential singularity.

\section{\label{sec:continuum}Continuum scaling with an action with strongly suppressed taste breaking} 

\begin{figure}[t]
  \centering
  \includegraphics[angle=270, width=0.98\linewidth]{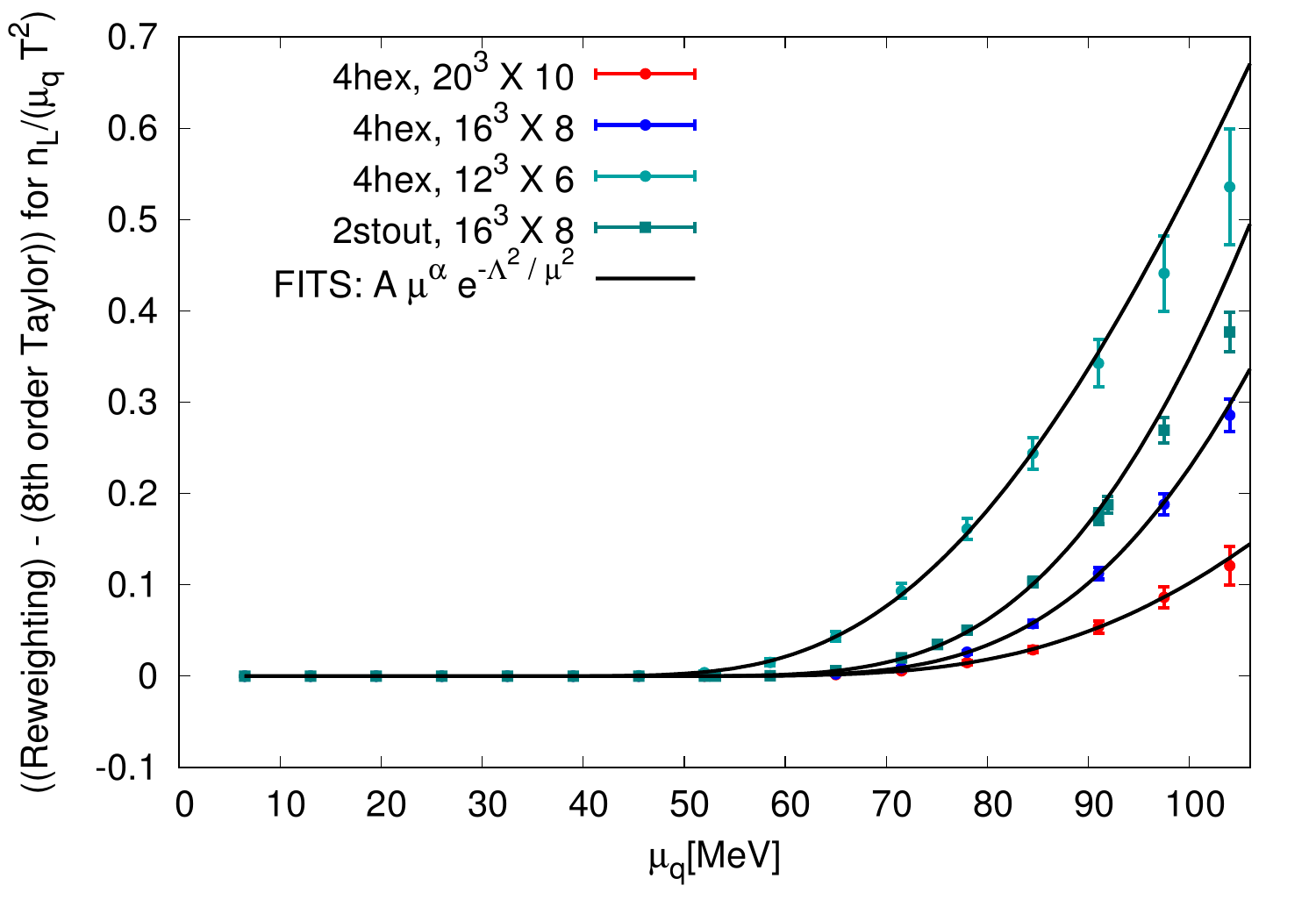}
  \caption{
      The difference between the full reweighted result and the 8th-order Taylor expansion
      (which is assumed to be a good proxy 
      for the analytic part of the density)
      for the light-quark density in units of the dimensionless quark chemical potential
       for the 4HEX action with 6,8 and 10 timeslices respectively. For comparison,
      results with the 2stout action with 8 timeslices are also shown. We also show fits of the form $A \mu_q^\alpha e^{-\Lambda^2/\mu_q^2}$. 
      The fits were performed in the chemical potential range up to $85$~MeV.
      \label{fig:subtract}
  }
\end{figure}

A natural question that arises is whether the observed non-analytic behavior of the free energy of rooted staggered fermions
vanishes in the continuum limit, or not. To study this question, we used a discretization with strongly suppressed taste breaking:
the DBW2 gauge action~\cite{Takaishi:1996xj,QCD-TARO:1999mox,DeGrand:2002vu} and 4 steps of hex 
smearing~\cite{Capitani:2006ni}, which we will call the 4HEX action~\cite{Borsanyi:2020mff}. For this simulation we used
physical quark masses~\cite{Borsanyi:2020mff} at temperature $T=130$~MeV, an aspect ratio $LT=2$, and three different 
lattice spacings corresponding to $N_\tau=6,8$ and $10$ time-slices each. 
In Fig.~\ref{fig:subtract} we show the difference between full reweighting and an 8th-order Taylor expansion for the ratio $\hat{n}_L/\hat{\mu}$.
For comparison, we 
also show results with the 2stout action at $N_\tau=8$. One can see that the 
magnitude of the difference decreases rapidly with the lattice spacing. 
We did not manage to find a good fit ansatz to extrapolate this difference to the continuum, however. In particular, we do not observe the $\mathcal{O}(a)$ scaling which is the expectation from the naive counting
argument given in~\cite{Golterman:2006rw}. This might be due to the coarsest lattice ($N_\tau=6$) being too coarse for seeing the asymptotic behavior. 
Notice, however, that if one assumes that the difference extrapolates to zero, then the observed decrease is not slower than the expected $\mathcal{O}(a)$, but 
faster: rescaling the $N_\tau=8$ data by a factor of $8/10$ gives numbers that are significantly above the lattice data at $N_{\tau}=10$. Even 
rescaling the $N_\tau=8$ data 
by $(8/10)^2$, the data on the $N_\tau=10$ lattices is significantly below. Comparison with the 2stout results shows that at the same number of timeslices,
the 4HEX action has a smaller non-analytic term. This is consistent with the expectation that reduced taste breaking (and thus the splitting of the taste multiplets in
the spectrum) reduces this non-physical effect in the quark density. 

\begin{figure}[t!]
  \centering
  \includegraphics[width=0.95\linewidth]{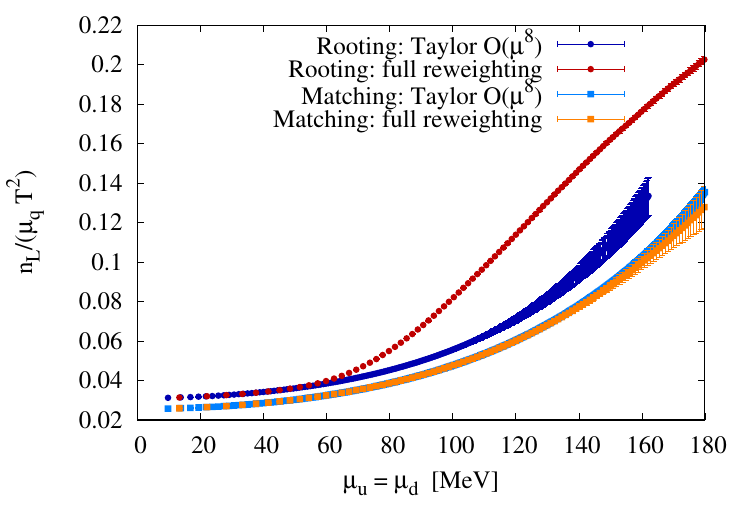}
  \caption{
Geometric matching vs standard rooting on the $12^3 \times 8$ 2stout ensemble
at a temperature $T=130$~MeV.  We show both full reweighting and Taylor
expansion results. The sharp rise around $\mu_q=m_\pi/2$ is not observed with 
the geometric matching
procedure.
\label{fig:matching}
  }
\end{figure}

\section{Summary and discussion}

We have shown that the free energy of QCD at nonzero baryochemical potential 
with rooted staggered fermions (with the rooting ambiguity resolved by requiring continuity for real values of the chemical 
potential configuration by configuration) has a non-analytic term, that manifests itself as a rapid rise of the light-quark density at low temperatures 
and baryochemical potential $\mu_B = 3\mu_q \gtrsim 3m_\pi/2$. This is at the same values of the quark chemical potential $\mu_q$ where the phase-quenched ensemble
has a pion condensation transition. We showed that the non-analyticity is very different from a true thermodynamic phase transition by comparing
with the case of an isospin chemical potential. 
Although the behavior of the light-quark density as a function of the chemical potential looks very similar in the two cases, the phase transition at nonzero isospin density leads to a branch-point singularity at some nonzero $\mu$, while rooting leads to an essential singularity at $\mu_B=0$.

How then can we proceed with the exploration of the QCD phase diagram at $\mu_B>0$?

The Taylor coefficients themselves can be interpreted as 
derivatives of the free energy with respect to an 
imaginary chemical potential, and thus can be defined 
without any ambiguous complex rooting. These can form a basis 
for an analytical continuation to nonzero $\mu_B$. However, 
the Taylor expansion does not define the theory non-perturbatively in
$\mu_B$. Furthermore, TODO

We have seen that the magnitude of the non-analytic part of the free energy
decreases with decreasing lattice spacings. With fine enough staggered lattices
one could attempt a continuum extrapolation, which could be free from this 
(presumably) unphysical non-analyticity. 
The use of very fine staggered lattices is, however, 
beyond feasibility today, especially if all eigenvalues 
of the reduced matrix are required to calculate the quark determinant.

An alternative approach, still
within the framework of staggered fermions, is offered by Ref.~\cite{Giordano:2019gev}. 
One can match and pair
the eigenvalues of the reduced matrix based on a geometric principle and replace
such pairs with a single eigenvalue, thus halving their number. Since
in the reduced matrix formalism the chemical potential is introduced after
the eigenvalues of the reduced matrix have been computed, one can loosely describe this
method as a way to do the rooting before introducing the chemical potential, 
rather than after.
In this definition of the determinant at nonzero $\mu_q$, non-analytic terms 
of the type
we empirically observed with standard rooting are explicitly forbidden, as the
free energy can be shown to be analytic at $\mu_B=0$~\cite{Giordano:2019gev}.
We demonstrate this in Fig.~\ref{fig:matching}, where we show the full
reweighting and Taylor expansion results for the geometrically matched
definition of the determinant. We observe good agreement between the
two procedures, and no rapid rise in
the light-quark density at $\mu_B \gtrsim 3 m_\pi/2$. This is in 
contrast to standard rooting which is also shown in Fig.\ref{fig:matching}.
(Note that the rooted and geometrically matched results 
were obtained using the same gauge ensemble.) While one
naively expects standard rooting and geometric matching to give the same
continuum limit, the confirmation of this expectation requires further
research. 

Note, however, that the geometric 
matching definition of the determinant is 
still unlike a true two-flavor formulation, as analyticity 
is guaranteed only in the chemical potential and not in the
gauge fields. In particular, while the eigenvalues of the 
reduced matrix are smooth functions of the gauge fields, 
a small change in the gauge fields can change which pairs are
identified in the spectrum of the reduced matrix, leading to
non-analyticity in the link variables. This is similar to
standard rooting, which is also not analytic in the fields, 
as changing the gauge fields can lead to crossing 
branch cuts. In both cases, this could randomize the 
phases of the quark determinant. For completely random phases,
uncorrelated with the observable of interest, one should
simply get phase-quenched physics:
\begin{equation}
\left\langle O \right\rangle = 
\frac{\left\langle O e^{i \theta} \right\rangle_{PQ}} {\left\langle e^{i \theta} \right\rangle_{PQ}} \underset{\rm{uncorr.}}{\approx} \frac{\left\langle O \right\rangle_{PQ} \left\langle e^{i \theta} \right\rangle_{PQ}} {\left\langle e^{i \theta} \right\rangle_{PQ}} = \left\langle O \right\rangle_{PQ}\rm{.}
\end{equation}
Even in less extreme cases, one still expects cut-off effects 
with staggered fermions to be in the direction that 
the ensemble at nonzero baryon density is closer to the 
phase quenched ensemble than it should be in the continuum.
This is a separate issue from analyticity in the 
chemical potential, and should also be present with the matching
definition. Indeed, for unimproved (and unsmeared) 
staggered fermions, 
reweighting results with rooting and geometric 
matching tend to be 
very close to each 
other~\cite{Giordano:2020uvk, Giordano:2020roi}.

Finally, the most obvious move in light of the observed non-analytical
feature of the rooted staggered result is to completely abandon the staggered
formulation. Wilson-type fermions offer single-flavor discretizations for 
fermions that can
be essential to define the quark determinants in settings where
nonzero baryon and isospin chemical potentials are needed simultaneously, 
like, e.g., in the core of neutron stars.
However, the explicitly broken chiral symmetry makes the Wilson formulation
less desirable in the vicinity of the chiral transition and in the
search of the critical end-point in the $T-\mu_B$ phase diagram.

Recently, two-flavor formulations with a $U(1)$ remnant of chiral symmetry, aka 
the minimally doubled fermions, have reemerged. 
As long as the two light flavors are degenerate this setup can be
used to define a rooting-free quark determinant at finite $\mu_B$.
For thermodynamic studies, the Karsten-Wilczek formulation 
\cite{Karsten:1981gd,Wilczek:1987kw} is a
natural choice. Its anisotropy is not a fundamental obstacle in 
thermodynamics where the temporal direction is already special. 
Its renormalization beyond the one loop level \cite{Capitani:2009yn} is 
subject to active research.

\section*{Acknowledgements}

The project was supported by the BMBF Grant
No. 05P21PXFCA. This work is also supported by the
MKW NRW under the funding code NW21-024-A. Further
funding was received from the DFG under the Project
No. 496127839. This work was also supported by the
Hungarian National Research, Development and Innovation
Office, NKFIH Grant No. KKP126769. 
This work was also supported by the NKFIH excellence 
grant TKP2021{\textunderscore}NKTA{\textunderscore}64.
The authors gratefully acknowledge the Gauss Centre for
Supercomputing e.V. (\url{www.gauss-centre.eu}) for funding
this project by providing computing time on the GCS
Supercomputers Jureca/Juwels~\cite{JUWELS} at Juelich Supercomputer
Centre, HAWK at Höchstleistungsrechenzentrum Stuttgart,
and SuperMUC at Leibniz Supercomputing Centre.

\providecommand{\noopsort}[1]{}\providecommand{\singleletter}[1]{#1}%

\end{document}